\documentclass[a4paper]{article}

\usepackage{INTERSPEECH2021}

\usepackage{amsmath,graphicx,bm}
\usepackage{amssymb,amsmath,epsfig, epstopdf, subfigure}
\usepackage{multirow}
\usepackage{xcolor}
\usepackage{graphicx}
\usepackage{booktabs}
\usepackage{mathtools}

\title{Sequence-level Confidence Classifier for ASR Utterance Accuracy and Application to Acoustic Models}
\name{Amber Afshan$^1$,Kshitiz Kumar$^2$, Jian Wu$^2$}
\address{
  $^1$Dept. of Electrical and Computer Engineering, University of California, Los Angeles, CA\\
  $^2$Microsoft Corporation, Redmond, WA}
\email{amberafshan@g.ucla.edu, \{kshitiz.kumar, jianwu\}@microsoft.com}

\begin{document}

\maketitle
\begin{abstract} 
Scores from traditional confidence classifiers (CCs) in automatic speech recognition (ASR) systems lack universal interpretation and vary with updates to the underlying confidence or acoustic models (AMs). In this work, we build interpretable confidence scores with an objective to closely align with ASR accuracy. We propose a new sequence-level CC with a richer context providing CC scores highly correlated with ASR accuracy and scores stable across CC updates. Hence, expanding CC applications. Recently, AM customization has gained traction with the widespread use of unified models. Conventional adaptation strategies that customize AM expect well-matched data for the target domain with gold-standard transcriptions. We propose a cost-effective method of using CC scores to select an optimal adaptation data set, where we maximize ASR gains from minimal data. We study data in various confidence ranges and optimally choose data for AM adaptation with KL-Divergence regularization. On the Microsoft voice search task, data selection for supervised adaptation using the sequence-level confidence scores achieves word error rate reduction (WERR) of 8.5\% for row-convolution LSTM (RC-LSTM) and 5.2\% for latency-controlled bidirectional LSTM (LC-BLSTM). In the semi-supervised case, with ASR hypotheses as labels, our method provides WERR of 5.9\% and 2.8\% for RC-LSTM and LC-BLSTM, respectively.
\end{abstract} 
\noindent\textbf{Index Terms}: Confidence Classifier, AM adaptation, AM customization,  Data selection, LSTM

\section{Introduction}\label{Sec:Introduction}
Deep learning techniques have led to highly accurate ASR systems, resulting in their widespread use. It is therefore critical to ensure the reliability of ASR hypotheses. A confidence classifier is one such integral part of an ASR system, providing a measure of ASR reliability. But the current confidence scores lack a common meaning and potentially change with variation to confidence model or AM. Hence, we develop a sequence-level confidence model with an objective to closely tie the CC score for an utterance to its ASR accuracy. Thus, providing interpretation and uniformity in confidence scores across updates to confidence model or AM and expand CC applications. We study one such use, AM customization. We propose a cost-effective method using confidence scores for optimal data selection for AM customization in supervised and semi-supervised cases.

ASR confidence applications include (a) mitigating false alarms in ASR-enabled devices in always-listening mode, (b) data selection for AM training and adaptation, (c) arbitration to select the best hypotheses between client and server results, (d) key-word spotting tasks. CCs are commonly built with multi-layer perceptron (MLP)~\cite{Huang2013a}, recurrent neural networks (RNNs)~\cite{Kalgaonkar2015EstimatingNetworks}, and bidirectional RNNs~\cite{Del-Agua2018Speaker-adaptedNetworks, Ragni2019ConfidenceNetworks}. These work at a word or a sub-word level~\cite{CMsurvey_Jiang_SpeechCommunication06}. Few methods use context~\cite{Sarma2004}, and alternative hypotheses to perform utterance verification~\cite{Setlur1996}. Recent works focus on training confidence models at an utterance-level to classify correct/incorrect utterances with applications to distributed end-to-end speech recognition systems~\cite{Kumar2020a, Li2021}. 

Conventional CC scores lack absolute interpretation and reflect the underlying CC training data, AM, language model, task, etc. So the scores from two different CCs are not directly comparable. Thus, needing to shift the confidence thresholds across CC updates or confidence normalization~\cite{Kumar2014} to recalibrate scores and retain those thresholds. While normalization preserves operating confidence thresholds across CC updates, it imparts no meaning to the scores. ASR developers or consumers expect CC scores to have a reasonable interpretation. We pursue that missing feature and introduce the ability to tie CC scores directly to ASR accuracy. In prior systems, a confidence score of say $0.9$ lacked an exact meaning except for the expectation that corresponding ASR hypotheses could be more accurate than the ones with lower confidence, say $0.8$. However, in the proposed CC, we expect a confidence score of $0.9$ for an utterance to imply 90\% accuracy. 

There is an increase in popularity of ASR systems~\cite{Ochiai2017UnifiedBeamforming, Fujita2016} built with ``unified data'', consolidated from domains like dictation, voice search, command-and-control, call center, etc.  The acoustic model trained on unified data--a ``unified ASR''--simplifies AM training, where a single large-scale model includes multiple acoustic scenarios. It also facilitates better data sharing and delivers higher accuracy on unseen tasks. Given sufficient data for a specific domain, it is cost-effective to customize the unified model to the target domain rather than training a domain-dependent model for higher accuracy~\cite{Li2020, Bell2020}.  

Change in the target domain, speakers, tasks, accents, etc., is dealt with using AM adaptation~\cite{Bell2020}. Typically, it needs data well-matched to the target with supervised annotations but is impractical due to privacy and data constraints. Previously, a seed model provided target labels in a semi-supervised manner, filtering data using normalized frame-level entropy~\cite{Liu2007}, and minimum Bayes risk~\cite{Walker2017}. Others~\cite{Manohar2018, Klejch2019} used multiple potential transcriptions with a lattice or graph. We aim to design an efficient and scalable approach to customize AM. We propose to use confidence scores for supervised and semi-supervised AM adaptation. If supervised, we optimally select data for subsequent transcription and adaptation with a goal to minimize transcription costs and maximize ASR adaptation gains. In semi-supervised cases, we chose higher-quality ASR hypotheses.

In this work, we propose a sequence-level LSTM-based confidence model whose scores align to ASR accuracy and stable across updates, eliminating the need for confidence normalization. Such a CC enables us to perform data selection and has scores tied to accuracy. We used CC scores to choose an optimal subset for AM adaptation developing a cost-effective method resulting in an improved WER. Besides, semi-supervised adaptation performance with optimal data is comparable to the supervised one with all the data. Section~\ref{sec:cc} presents the proposed sequence-level CC. Section~\ref{sec:am} details our recommendations for optimal data selection to customize AM. The experimental setup are in Section~\ref{sec:exp} and the results and discussion in Section~\ref{sec:results}. We conclude with Section~\ref{sec:conclusion}.
\section{Sequence-Level Confidence Classifier to Predict ASR Accuracy}
\label{sec:cc}
Confidence classifiers predict the correctness of the ASR hypotheses. ASR errors can include substitution, deletion, insertion, and potential hypotheses over non-speech sounds and background noise. CC training used a set of features described in~\cite{Huang2013a}, which typically includes acoustic model score, language model score, word duration, and count of phonemes. 
\subsection{MLP Confidence Classifier}
\label{sec:MLP}
Our baseline is a word-level MLP classifier. The high accuracy of ASR systems results in imbalanced correct and incorrect words with fewer negative samples, requiring data balancing across the two classes. We trained baseline with cross-entropy criterion providing word-level confidence scores. CC applications, including AM adaptation addressed in this work, require confidence scores at an utterance level. Since an unweighted aggregation has shown to be less effective than a weighted average of the word-level scores with corresponding word duration~\cite{CMsurvey_Jiang_SpeechCommunication06}, we obtain utterance-level scores from a duration weighted average. Two main drawbacks of MLP CC are that every word is processed independently of the rest of the utterance, and the aggregation of the utterance-level information happens at the end, providing confidence scores with no interpretable meaning. A sequence model can explicitly process information at an utterance level and directly produce confidence scores per utterance. 
\subsection{LSTM Sequence Confidence Classifier \label{sec:lstm}} 
We noted that the traditional confidence scores lacked meaning, requiring operating threshold adjustments across CC or AM model updates. Confidence normalization can map the scores and retain thresholds, but it is still desirable to impart an interpretable meaning to scores. We seek a close parity between accuracy and CC scores. To better capture accuracy, we need to process data at an utterance level, motivating a sequence-level CC. It is well-known that ASR models (LSTM~\cite{Sak2014}, TDNN~\cite{Povey2016}) benefit from sequence information. We expect such benefits to extend to the CC model. Hence, we propose a sequence-level confidence model using LSTM trained on word-level features, performing a regression task with the cumulative sum of errors as labels. At a given time instant, LSTM CC predicts a CC score on the hypothesis from the beginning of the utterance until that instant. It generalizes, and the final prediction of the utterance reflects the confidence for the entire utterance. The cumulative errors as target labels in our design resulted in a dependency on utterance length. To maintain uniformity, we apply length normalization to the cumulative errors. The training loss is,
\vspace{-2em}
\begin{align}
    L &= \sum_{n=1}^N l_n, \hspace{1em} l_n = (\hat{y}_n - y_n)^2 \\
    y_n &= \frac{\# \textrm{ correct words in hypothesis}}{n}
\end{align}

\noindent where $l_n$, $y_n$, $\hat{y}_n$  are loss, accuracy (1-WER), and prediction, respectively at the $n^{th}$ word of $N$-word utterance. $\hat{y}_N$ is the utterance confidence score. Although the deleted words are not assigned CC scores, the accuracy calculations include deletion errors. We expect our optimization to provide utterance-level CC scores tightly correlated to accuracy, as LSTMs capture context and embed sequential information in the CC. While we chose LSTM as the confidence model, we expect this work to generalize to other sequence-level models like Transformers.
\begin{table*}[t]
\begin{center}
\caption{{\it Recommendation for selecting optimal data for AM adaptation based on the range of confidence scores. We select a percentage of data as shown in examples instead of setting an operational threshold. }}\label{table:selection}
\resizebox{0.95\linewidth}{!}{%
\begin{tabular}{lll|cc|l}
\toprule
\multicolumn{1}{c}{\multirow{2}{*}{\textbf{Confidence Range}}} &
  \multirow{2}{*}{\textbf{Examples}} &
  \multicolumn{1}{c|}{\multirow{2}{*}{\textbf{\begin{tabular}[c]{@{}c@{}}Accuracy\\ (Predicted)\end{tabular}}}} &
  \multicolumn{2}{c|}{\textbf{AM Adaptation Recommendation}} & \multirow{2}{*}{\textbf{Comments}} \\
  \cmidrule{4-5}

\multicolumn{1}{c}{} &          & \multicolumn{1}{c|}{} & \textbf{Supervised} & \textbf{Semi-supervised} & \\
\midrule
\midrule
High                 & Top 20\% & High                 & -                  & \checkmark$^+$                & Can be proxy for manual transcriptions.\\
Mid                  & Top 20-40\%, Top 40-60\%  & Partially accurate        & \checkmark                  & \checkmark                  &  Can provide partial improvement.    \\
Penultimate          & Bottom 10-30\%, Bottom 20-40\%  & Low                  & \checkmark $^+$          & -                &   Can learn from manual transcriptions.   \\
Very low                  & Bottom 0-10\%  & Lowest   & -                  & -      &  Nuisances and unwanted recordings.\\
\bottomrule

\multicolumn{6}{p{\linewidth}} {$^+$ The subset that provided the maximal gain with minimal data. }\\
\vspace{-5em}
\end{tabular}%

}
\end{center}
\end{table*}
\subsection{Context Information for Confidence Classifier}
Context is another source of information for a better CC. It can learn and distill valuable knowledge from the inter-dependencies within an utterance (for example, co-articulation). We propose stacking the features across a window of a few past and future words to capture the context in  CC models. This simple approach to including context could benefit all CC models.
\section{Optimal Data Selection for Acoustic Model Customization }
\label{sec:am}
AM training at a commercial scale is an intensive process, and unified AM eliminates the need to train and deploy AM for individual scenarios. Customization is a cost-effective way to incorporate domain-dependent data in the unified model to obtain higher gains on specific domains. Model customization is especially valuable for large-scale customers who want to focus the unified model on particular application scenarios. Customization is more efficient than adding domain-specific data to the original training pipeline or developing a domain-dependent model from scratch. Customization can leverage complex adaptation approaches given the availability of an adequate amount of data. A scalable, lower-cost solution for AM customization is desirable that can potentially work with training data at different scales. We apply customization to two unified AM baselines: LC-BLSTM~\cite{Xue2017,Chen2016} and RC-LSTM~\cite{Kumar2020}. 
\subsection{Acoustic Model Structures}
The BLSTMs learn from both past and future information, providing gains over LSTMs but at the cost of increased complexity, model size, and latency. LC-BLSTM addresses some of these issues by restricting the forward and backward computation to batches of data with minimal loss in ASR accuracy. RC-LSTM uses a row convolution structure to distill and embed some future information into uni-directional LSTMs. It implicitly factorizes training in orthogonal time and ``frequency" components leading to efficient training for large-scale models. RC models have smaller lookahead and are computationally efficient, but LC models access additional future information and are more complex with higher accuracy than RC-LSTM.
\subsection{KL-Divergence Adaptation}
The training data for model adaptation is often limited. We, therefore, build a conservative adaptation approach. We choose from regularized adaptation techniques: $L2$ regularized loss~\cite{liao2013speaker}, KL-divergence (KLD) regularized loss~\cite{YuKL2013}, etc. We use KLD adaptation where KLD measures the distance between distributions of original and adapted models with parameters $\theta$ and $\theta'$, respectively. Given acoustic features $X= (x_1,x_2,\dots,x_T)$ of length $T$ the loss is,
\vspace{-0.5em}
\begin{equation}
    \mathcal{L}_{KL} = \mathcal{D}_{KL}(f(x;\theta)||f(x;\theta'))
\end{equation}
where $f()$ is the AM with parameters $\theta$ outputting labels $Y=(y_1,y_2,\dots,y_T)$. 
We used the KL distance along with cross-entropy training criteria ($\mathcal{L}_{xent}$), resulting in an overall loss,
\vspace{-0.5em}
\begin{equation}
    \mathcal{L} = (1-\lambda)\mathcal{L}_{xent} + \lambda \mathcal{L}_{KL}
\end{equation}
$\lambda$ is the weight of the original model and $\hat{P}(Y|X)$ is the distribution of provided labels of the adaptation data $y^{adapt}$. The loss is equivalent to cross-entropy criteria with target distribution:
\begin{equation}
\label{eqn:distribution}
    P(Y|X) = (1-\lambda)\hat{P}(Y|X) + \lambda f(x;\theta)
\end{equation}
\subsection{Optimization of Data Used in AM Customization}
\label{sec:optimization}
Budget constraints and scalability needs bound any data collection and transcription for custom AMs. There are constant changes in data, systems, ASR engines, etc. Finally, not all data is equal in providing ASR gains. Hence, we develop a compact adaptation set by optimally selecting data for AM customization. We seek minimal data that yields maximum gain. In supervised adaptation, the primary cost is the transcription of the customization data. So in a supervised case, where transcriptions for the data are labels, we intend to optimally select a minimal set of data for subsequent transcription and adaptation. In semi-supervised adaptation, where ASR hypotheses are labels for the data, we choose ASR hypotheses with high accuracy to form the adaptation set. In both cases, confidence scores are crucial in selecting optimal data for cost-effective customization.

Our recommendations are in Table~\ref{table:selection} for using confidence scores to select data in supervised and semi-supervised methods. We have a trade-off between the amount of data and expected data quality, so we consider confidences and bucket the data in percentile bins. We recommend high confidence data for semi-supervised adaptation as we expect unified ASR to be highly accurate in this case. However, we chose penultimate confidence range data for supervised adaptation as unified ASR is already accurate at high and mid confidence bins. The above follows that there is more scope to learn from penultimate confidence data. We exclude very-low confidence scores because that data likely has nuisances like highly overlapped speech, no speech, speech not targeted at the device, etc. The mid-range confidence scores can be appropriate for either approaches.
\section{Experimental Setup}
\label{sec:exp}
\begin{table}[t]
\begin{scriptsize}
\begin{center}
\caption{{\it Statistics of the database used for the acoustic model.}}
\label{table:Data}
\begin{tabular}{lc}
\toprule
Database & Number of hours\\
\midrule
\midrule
en-US Training & 64,000 \\
Voice search Adaptation  & 35 \\ 
Voice search Test & 3\\ 
\bottomrule
\vspace{-2em}
\end{tabular}
\end{center}
\end{scriptsize}
\end{table}

We used a large-vocabulary en-US data for unified AM training. We trained the confidence model on $\sim$60k~utterances sampled from the en-US data, along with $\sim$7k~utterances for the CC test set. A universal data set, the en-US data comprised of data from tasks such as voice search, close-talk, natural conversation, command-and-control, dictation,  call center, etc. We customized the unified model to a voice search task. This voice search task used for customization is from a different target application and is more recent, with queries from the latest trends than the en-US data. Table~\ref{table:Data} has the statistics of the adaptation data. All the training and test data were anonymized, with all personal identifiable information removed. Our proposed CC uses a 4-layer unidirectional LSTM model with 512 cells with MSE training criteria. We removed one-word recordings from both training and testing lists of the LSTM. LSTM embeds and builds context, but the one-word utterances offer no context, and we found better convergence by excluding these utterances. We extracted 80-dim log-Mel features from 25~ms frames and a 10~ms frame-shift for ASR and used a 5-gram language model with a vocabulary of over 1M words. Adaptation was using KLD regularization with a KL weight, $\lambda=0.5$.
\section{Results and Discussion}
\label{sec:results}
\begin{figure}[t]
    \centering
    \vspace{-1em}
    \includegraphics[scale=0.22, trim={1.2cm 0.3cm 2cm 1.2cm}, clip]{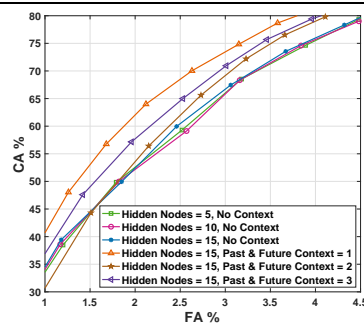}
    \caption{MLP CCs compared using CA-FA curves.}
    \label{fig:mlp_results}
    \vspace{-1em}
\end{figure}

\begin{figure}[t]
    \centering
    \includegraphics[scale=0.22, trim={1.5cm 0.15cm 2.5cm 1.3cm}, clip]{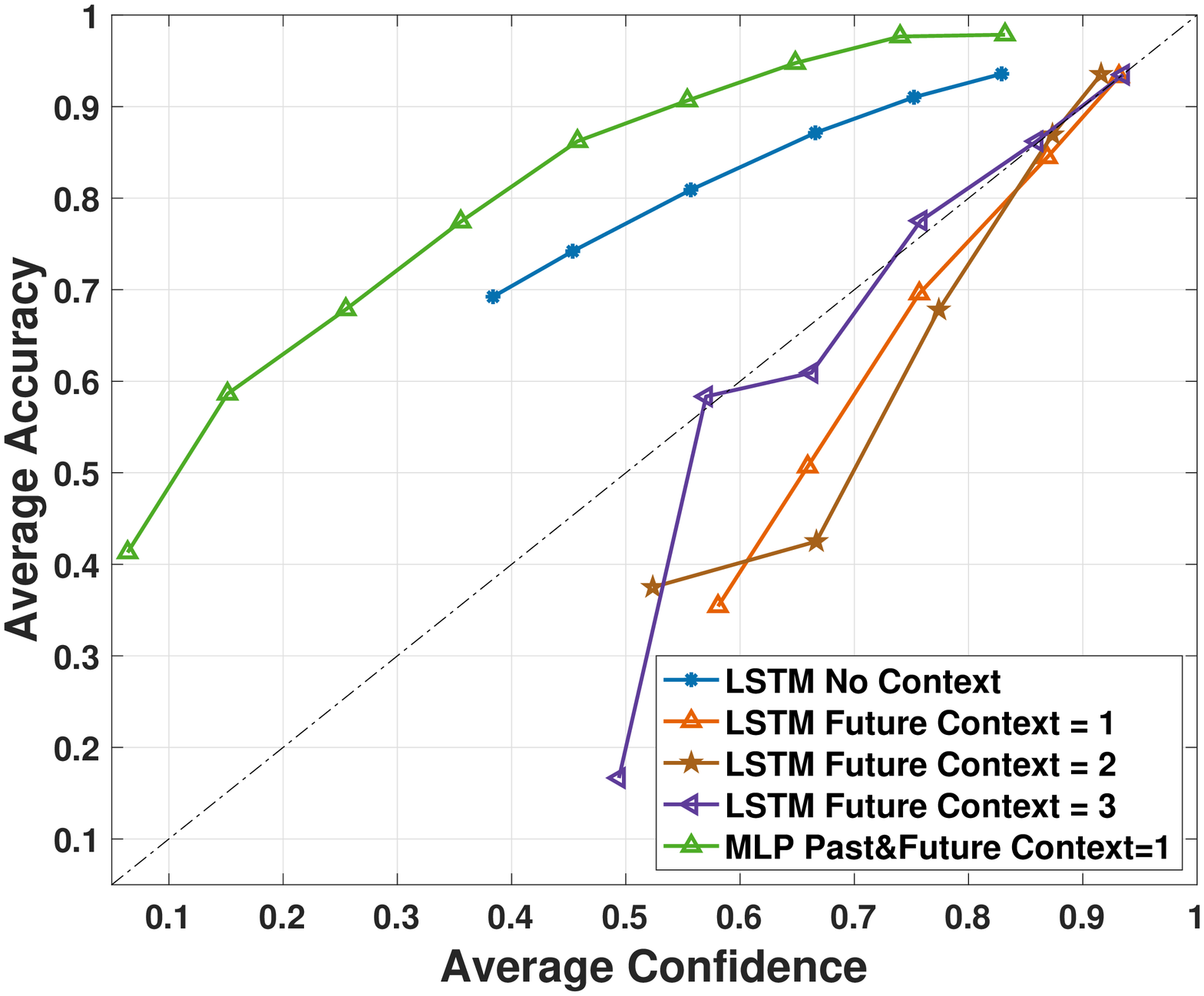}
    \caption{MLP compared with LSTM CC and context addition.}
    \label{fig:lstm_results}
    \vspace{-2em}
\end{figure}

\subsection{Confidence Classifier}

\begin{table}[b]
\begin{scriptsize}
\begin{center}
\vspace{-2em}
\caption{Comparison of the correlations (average bin accuracy and confidence scores) for tests on en-US and voice search data.} \label{tab:correlation}
\begin{tabular}{ccc}
\toprule
\textbf{Test data}            & \textbf{CC (train: en-US)} & \textbf{Correlation} \\
\midrule
\midrule
en-US        & LSTM        & 0.99            \\
     (matched)                         & MLP         & 0.95            \\
                              \midrule
Voice search & LSTM        & 0.97            \\
            (mis-matched)                  & MLP         & 0.78           \\
\bottomrule
\vspace{-3em}
\end{tabular}%
\end{center}
\end{scriptsize}
\end{table}
Figure~\ref{fig:mlp_results} studies the number of hidden nodes and the inclusion of context to the baseline MLP CC. We measure performance as correct acceptance (CA) and false acceptance (FA) rates, where we desire an operating range with high CA and low FA rates. 
\begin{equation*}
\scriptsize{
   CA = \frac{\#\text{ Corrects beyond an threshold}}{\#\text{All corrects}},
FA = \frac{\#\text{Incorrects beyond a threshold}}{\#\text{All incorrects}}  
}
\end{equation*}
In MLP CC, more neurons led to negligible performance gain, likely due to limited variability to encapsulate at word-level. But adding past and future contexts improves performance. We obtain the best performance with the context of one word. 

CA-FA curve, commonly used with word-level CCs, is an inadequate measure of performance for a CC model trained at utterance-level in a regression setting. Besides, we aim to align confidences with ASR accuracy (1-WER), so we devise a new metric for the LSTM CCs in Figure~\ref{fig:lstm_results}. We preserve the core idea in CA-FA curves and divide the confidences into ten bins at $0.1$ intervals. We produce average confidence and associated accuracy of utterances in those bins. Note that accuracy calculations include deleted words but they were not assigned confidence. We seek a high correlation between ASR accuracy and confidence, i.e., curves closer to the line of equality. 

Figure~\ref{fig:lstm_results} shows performance comparison for MLP and LSTM CCs. The confidence versus accuracy plot is closer to the line of equality for LSTM than MLP CC and concludes that LSTM CCs better predicts ASR accuracy. Confidence scores are better correlated to accuracy in higher confidence ranges demonstrating the effectiveness of sequential modeling of confidence scores. We show the benefit of adding future context in LSTM CCs with a few irregularities in the low confidence regions. We believe that is due to shorter utterances with limited future context that may dominate the low confidence regions. Additional future context may help LSTM CC for longer utterances but will have challenges due to limited training data and may regress on shorter utterances. We found a good trade-off with a future context of one word. 

Earlier analysis used the en-US train and test data. We examine the performance of data mismatch by evaluating the voice search task in Table~\ref{tab:correlation} for CCs trained with en-US. The results are in terms of correlation between the bin accuracies and confidence scores. As hypothesized, the proposed LSTM CC suffers little degradation compared to baseline MLP CC with data-mismatch between train and test. Thus, implying confidence normalization is unnecessary for the LSTM CC, making it more suitable for AM customization to different domains. 

\subsection{Customization of Acoustic Models}


\begin{table}[t]
\centering
\caption{{\it {Word error rate (WER) and WER reduction (WERR) over the baseline unified model for customization to a voice search task using different optimal subsets (see Table~\ref{table:selection}).}}}\label{table:adaptation}
\resizebox{0.95\linewidth}{!}{%
\begin{tabular}{c|c|cc|cc}
\toprule
        &  \multirow{2}{*}{\textbf{Range}}     & \multicolumn{2}{c}{\textbf{RC-LSTM}} & \multicolumn{2}{c}{\textbf{LC-BLSTM}} \\
\cmidrule{3-6}
                            &      & WER\%           & WERR\%        & WER\%           & WERR\%         \\
                            \midrule
                            \midrule
Baseline                    &      & 13.61         & -           & 13.25         & -            \\
       \midrule
\multirow{3}{*}{Sup} & All  & 12.74         & 6.39        & 12.85         & 3.02         \\
                            & Top 20\%    & 12.72         & 6.54        & 12.88         & 2.79         \\
                            & Bottom 10-30\%    & \textbf{12.46}         & \textbf{8.45}      & \textbf{12.56}         & \textbf{5.21 }        \\
                                   \midrule
\multirow{3}{*}{Semi-sup}            & All  & 13.28         & 2.42        & 13.02         & 1.74         \\
                            & Top 20\%  & \textbf{12.81 }        &\textbf{ 5.88 }       & \textbf{12.88}         & \textbf{2.79 }        \\
                            & Bottom 10-30\%    & 12.99         & 4.56        & 13.03         & 1.66   \\
                            \bottomrule
\end{tabular}%
}
\vspace{-2em}
\end{table}

We customize unified AM to a voice search task and report WER and WER reduction (WERR) over baseline unified AM in Table~\ref{table:adaptation}. Model structures used are RC-LSTM and LC-BLSTM as described in~\cite{Kumar2020}. The LC-BLSTM unified AM, as expected, performs better than RC-LSTM.

Supervised adaptation with all the adaptation data (35~hours) shows 6.39\% WERR for RC and 3.02\% WERR for LC model. A relatively higher gain for the RC model shows that simple model structures can be more effective for limited data adaptation. These results provide a reference for subsequent AM customization following our recommendations in Table~\ref{table:selection} of optimal data selection with confidence scores. As hypothesized, supervised adaptation with penultimate confidence range data (7~hours) performs the best with WERRs of 8.45\% and 5.21\% for RC and LC models, respectively. Thus, saving transcription costs significantly by transcribing data only from the penultimate range. Semi-supervised adaptation performed best using a higher confidence range data, the top 20\% of the data (7~hours), resulting in WERR of 5.88\% and 2.79\% for RC and LC models, respectively.  CC can hence effectively select data with reliable hypotheses for semi-supervised adaptation. Finally, our optimal data selection is better than all data in both cases. Our reported evaluations are on large-scale tests, and we expect any WER beyond 1\% relative to be significant. Note that we do not report the mid and very-low range as they are not optimal and result in very low WERR. As an extension, we also combined supervised (ground-truth labels; bottom 10-30\%) and semi-supervised (ASR hypothesis as labels; top 20\%), a total of 14~hours of data to customize the LC model. We obtained a WERR of 4.08\%, not outperforming the corresponding supervised-only adaptation. We also conducted similar experiments with confidences from MLP CC, resulting in some improvement over the baseline. For instance, the top 20\% data selected based on MLP CC scores resulted in a WERR of 5\% in a supervised adaptation.  
\section{Conclusion}
\label{sec:conclusion}
We proposed a sequence-level LSTM-based confidence model and captured a rich context, resulting in confidence scores that closely followed accuracy--more prominently in high accuracy brackets. Traditionally, confidence scores lack universal interpretation that changes across CC updates. The proposed CC introduced scores that depict the accuracy and eliminate the need for confidence normalization across CC updates, making them suitable for many tasks. The CC, therefore, enabled the use of confidences for adaptation data selection by providing stable utterance-level scores that predict accuracy.  Consequently, we developed an efficient, cost-effective, and scalable approach for AM customization using CC scores to select optimal adaptation data. This work is the first evidence that compact data set selected by a CC is better than using all data--both in supervised and semi-supervised AM customization. In the proposed approach, we chose an optimal subset of the data to perform transcription and reduce cost while obtaining maximal gain. Likewise, we chose a subset of optimal data with high-quality ASR hypotheses to act as labels. Semi-supervised adaptation with optimal data resulted in WERR comparable to the supervised one with all the data, proving the efficacy of the proposed method.  Given the compact nature of the RC model, it had more benefits from customization when compared to the LC model.  
\bibliographystyle{IEEEtran}

\bibliography{references}

\end{document}